\documentclass[conference]{IEEEtran}
\IEEEoverridecommandlockouts
\usepackage{cite,url}
\usepackage{amsmath,amssymb,amsfonts}
\usepackage{algorithmic}
\usepackage{graphicx}
\usepackage{textcomp}
\usepackage{xcolor}

\newcommand{\fref}[1]{Fig.~\ref{#1}}

\def\BibTeX{{\rm B\kern-.05em{\sc i\kern-.025em b}\kern-.08em
    T\kern-.1667em\lower.7ex\hbox{E}\kern-.125emX}}
\begin{document}

\title{A System of Bidirectional Power Routing Toward Multi-energy Management *\\
\thanks{This work was partially supported by JSPS KAKENHI Grant Number JP24K17262. }
}

\author{\IEEEauthorblockN{Shiu Mochiyama}
\IEEEauthorblockA{\textit{Department of Electrical Engineering,} \\
\textit{Kyoto University}\\
Kyoto, Japan \\
s-mochiyama@dove.kuee.kyoto-u.ac.jp}
\and
\IEEEauthorblockN{Ryo Takahashi}
\IEEEauthorblockA{\textit{Department of Mechanical and Electrical Systems Engineering,} \\
\textit{Kyoto University of Advanced Science}\\
Kyoto, Japan}
\and
\IEEEauthorblockN{Yoshihiko Susuki}
\IEEEauthorblockA{\textit{Department of Electrical Engineering,} \\
\textit{Kyoto University}\\
Kyoto, Japan}
\and
\IEEEauthorblockN{Tsutomu Wakabayashi}
\IEEEauthorblockA{\textit{Energy Technology Laboratories,} \\
\textit{Osaka Gas Co., Ltd.}\\
Osaka, Japan}
\and
\IEEEauthorblockN{Takumi Tanaka}
\IEEEauthorblockA{\textit{Energy Technology Laboratories,} \\
\textit{Osaka Gas Co., Ltd.}\\
Osaka, Japan}
\and
\IEEEauthorblockN{Toshinari Momose}
\IEEEauthorblockA{\textit{Residential Energy System Development Department,} \\
\textit{Osaka Gas Marketing Co., Ltd.}\\
Osaka, Japan}
\and
\IEEEauthorblockN{Hideki Yamaguchi}
\IEEEauthorblockA{\textit{Residential Energy System Development Department,} \\
\textit{Osaka Gas Marketing Co., Ltd.}\\
Osaka, Japan}
}

\maketitle

\begin{abstract}
In this paper, we propose a system of bidirectional power routing for inter-house multi-energy management systems that utilize electricity and hydrogen as energy carriers. 
The key is to share private facilities such as photovoltaic panels and batteries among a group of houses along with a common hydrogen system. 
A power router of line switching type is introduced as a physical interface to realize the sharing economy between households. 
The proposed system offers a unique measure to address the urgent challenges of today's multi-energy system, namely increasing the renewables' self-consumption, enhancing the energy system's resilience, and providing traceability of hydrogen in terms of renewability certification. 
We also present an experimental demonstration under a simplified scenario using prototype hardware.
\end{abstract}


\section{Introduction}

Energy systems of multiple energy carriers have attracted much attention recently. 
Several concepts have been proposed around this topic, including multi-carrier energy systems \cite{Geidl.Andersson-2007,Geidl.etal-2007}, integrated energy systems \cite{OMalley.Kroposki-2013,Wu.etal-2016} and multi-energy systems \cite{Mancarella-2014,Chertkov.Andersson-2020}. 
The energy carriers in such systems are heat, natural gas, hydrogen, and so on, as well as electricity. 
The interactions between these carriers, which were traditionally handled rather independently, are expanding according to the rapid development of their coordination and mutual conversion technologies, e.g., applications of combined heat and power units \cite{Ostergaard-2010} to frequency regulation of electricity grids \cite{Hoshino.etal-2023}. 

The present study focuses on the multi-energy systems of electricity and hydrogen. 
The key conversion technologies are fuel cell (FC) and power-to-gas (P2G) \cite{Staffell.etal-2019}. 
Due to its larger energy density than chemical materials used in electrical batteries\cite{Evans.etal-2012}, hydrogen provides energy management with unique features such as inter-seasonal energy storage. 

This study aims to develop a multi-energy management system using FC and P2G for a local community, such as an apartment complex and a group of individual houses. 
Each house is supposed to be equipped with photovoltaic (PV) generation and stationary and/or vehicle-mount batteries. 
In addition, we assume the community has P2G and FC facilities. 
The energy systems we cover in this study are facing urgent challenges related to including renewable supplies. 
Below, we describe the challenges addressed in our proposal. 

\subsection{Challenge I: Increasing Self-consumption of Renewables}

The demand for sustainability has driven the introduction of household PVs under substantial government subsidies. 
This situation, however, poses a challenge to the existing commercial grid\cite{Liang-2017}. 
The output of PV depends largely on weather conditions and does not always match the power consumption profile of the panel owner's house. 
Although an option is the reverse flow of excess power to the grid, the unpredictable fluctuation hinders the stable operation of the commercial grid. 
After all, reverse flow is often strongly restricted, leading to limited utilization of PV.
There needs to be another way to utilize redundant electricity from one house in another that requires it. 

Furthermore, the financial benefit of each household is also unignorable. 
Considering the cost of PV installation, we need to motivate the houses in terms of monetary advantage. 
Particularly, for the sustainable penetration of PV, there needs to be a motivation apart from governmental subsidies. 

A key technology to these challenges is the usage of sharing economy\cite{Sundararajan-2016}, namely benefiting from sharing private properties such as PV panels and batteries among houses. 
In fact, the impact of this concept in the area of electricity was suggested in \cite{Kalathil.etal-2019}, which reported that each house involved can benefit from sharing their residential electricity storage compared to the case where each buys and sells with a retailer on its own. 
Still, a remaining challenge in \cite{Kalathil.etal-2019} is the physical implementation of the sharing system to realize transactions of electricity. 

\subsection{Challenge II: Boosting up Resilience}

Another urgent challenge is the resilience of the power supply. 
There are risks of many types in power supply today, including the seasonal variation of demand due to climate change, political issues in fuel procurement, and aged facilities of the grid. 
For example, following the severe weather forecast, the Japanese government requested to save electricity multiple times in recent years\cite{TheJapanTimes-2023}. 
It is vital to depend less on the external grid for residential power availability. 

Furthermore, in an emergency where, e.g., the external supply is completely lost, there might be an extreme supply shortage compared to the expected load consumption. 
In such cases, the loads should be supplied according to their priority. 
Unfortunately, the current power system does not cover this operation except for the request for voluntary actions.

\subsection{Challenge III: Tracing Electricity for Hydrogen Production}

The climate-friendliness of hydrogen as an energy carrier largely depends on how it is produced. 
For example, hydrogen produced from fossil fuels is cheaper and constitutes a large part of today's total production \cite{Staffell.etal-2019}, but it yields unignorable carbon emissions. 
To achieve carbon neutrality, it is required to enhance the proportion of renewable hydrogen \cite{Gregor.Svensson-2023}, which is produced through electrolysis with renewable electricity. 

Now, the challenge occurs when electricity for electrolysis comes from the grid. 
Grid power usually originates from both renewable and fuel-based generations. 
It is difficult to physically distinguish how much of the generated hydrogen is renewable after production. 
Power flows from a renewable supply and fuel-based or mixed generations must be distinguished to ensure the origin of hydrogen; that is, their traceability is of technological concern. 

\subsection{Contributions of This Paper}

This paper proposes a system of power routing to physically realize the aforementioned multi-energy management system in a local community. 
Power routing is a technology for controlling power flows among multiple supplies and demands \cite{Takuno.etal-2010}. 
The power router in the proposed system serves as a physical interface to form the sharing economy of private facilities apart from the commercial grid. 
The physical device, as the key enabler of the concept on the hardware aspect, makes it feasible for the first time to trace transactions of electricity among the houses and the community-owned facilities, which offers a unified solution to the above three challenges in the current energy management systems. 
It should be noted that our proposal is along with physical system development. 
Although a sharing economy in the energy sector was studied in \cite{Kalathil.etal-2019}, its physical implementation has not yet been established. 
In this paper, we present a physically viable system of power routing based on a hardware technology.
We then show an experimental verification to support the viability of our proposal by demonstrating the proposed power routing in a simplified scenario with a prototype of the power router.

\section{Literature Review}

The concept of power routing dates back to the 1990s, the proposal by Toyoda and Saitoh \cite{Toyoda.Saitoh-1998}. 
They proposed an open electric energy network for distributed generations to participate in the electricity market. 
An energy router was proposed to serve as a hub for energy transactions with the support of information and communication technology. 
Unfortunately, the concept did not reach real implementation due to the immature hardware and software technologies at the time. 
After that, the development of the related technologies brought attention to the concept of power routing again in the early 2010s.
Many proposals were made by independent groups, e.g., power router \cite{Takuno.etal-2010}, energy router \cite{Huang.etal-2011}, smart power router \cite{Nguyen.etal-2011a}, digital grid router \cite{Abe.etal-2011}, smart green power node \cite{Stalling.etal-2012a}, and Y-configuration power router \cite{Kado.etal-2016a}.
Although the above proposals are mainly for residential power distribution systems, similar concepts have recently been expanded to closed and application-specific sectors such as transportation powering systems \cite{Yu.etal-2023,Liu.etal-2024}. 

From the viewpoint of physical implementation of power routers, these proposals can be divided into two categories. 

One is the line-switching type proposed in \cite{Takuno.etal-2010}. 
The circuit comprises semiconductor switches that are mutually connected in a matrix configuration. 
The circuit operates as a crossbar switch of power flow. 
The connection relationship between pairs of the circuit ports is determined by the switching states. 
Power routers have a communication function through power line communication to determine its switching table in response to the source and load requirements. 

In \cite{Takuno.etal-2010}, a power router of line-switching type was introduced to realize the concept of in-home power routing, aiming to increase the penetration of renewable generations while avoiding complexity and losses due to the requirement of high-quality power conversion for interconnecting to the grid. 
The idea originates from the variety of home appliances in terms of power quality requirements. 
For example, battery-powered portable devices such as laptops have tolerance to interruptions in power supply, while medical devices must not be interrupted. 
To achieve the best match between the various loads, renewable supplies, and utility power, the proposal adopted the line-switching type for the circuit configuration. 
The proposal was followed by the experimental verification presented in \cite{Takuno.etal-2011b}, where the router operation was demonstrated under the setup of two source-load pairs. 

Besides the line-switching, another type of implementation is based on multi-stage power conversion \cite{Huang.etal-2011,Abe.etal-2011,Stalling.etal-2012a,Kado.etal-2016a,Yu.etal-2023,Liu.etal-2024}. 
They share the basic concept that a router comprises multiple power conversion elements although there is a little difference in their detailed implementation.
In \cite{Huang.etal-2011,Kado.etal-2016a,Yu.etal-2023}, the hardware of power routers comprises power converters and a high-frequency AC coupling transformer. 
The in- and out-flows of power are put together and distributed through the common magnetic core. 
In \cite{Liu.etal-2024} a similar configuration is adopted where the high-frequency coupling transformer is replaced with a wireless power transfer system to accommodate both wired and wireless power flow. 
In \cite{Nguyen.etal-2011a,Abe.etal-2011}, the router circuit is back-to-back AC-to-DC and DC-to-AC converters. 
Input and output power of the router's ports are regulated by controlling the in- and out-flows of a common DC bus through AC-to-DC and DC-to-AC power conversion elements. 
In \cite{Stalling.etal-2012a}, the router is composed of bidirectional DC-to-DC converters and bidirectional inverters connected through a common DC bus. 

An advantage of power routers of the power conversion type is the capability to handle different voltage levels and/or a mixture of AC and DC. 
That is, the router ports can be connected to kV-order distribution lines, 120/240V loads, and DC supplies and loads. 
By taking advantage of this feature, there is a report on the application of a power router to a multi-energy system of hydrogen and electricity in \cite{Yu.etal-2023}. 
However, the power router of this type does not fully fit one of our targets, namely the power flow distinguishment with traceability of the origins. 
This is why we adopt the line-switching type for power router implementation in this paper. 


\section{Proposed Power Routing System}

\begin{figure}[tbp]
    \centering
    \includegraphics[width=6cm]{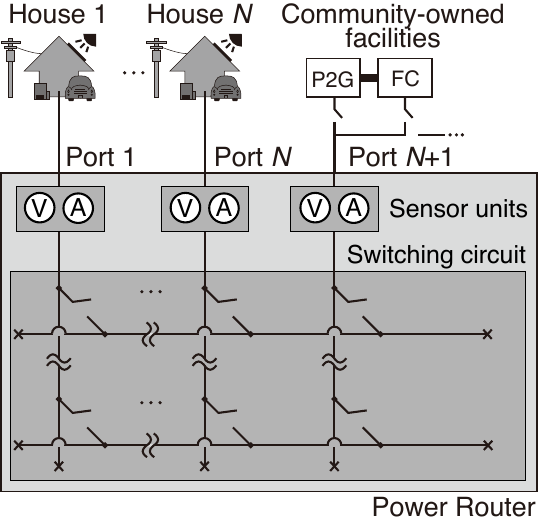}
    \caption{Example of the energy system for a group of $N$ houses and community-owned facilities. }
    \label{fig:system_concept}
\end{figure}

\fref{fig:system_concept} depicts an example of the energy system considered in this study. 
We assume a system of $N$ houses and community-owned facilities. 
Each house consists of loads, PV, and batteries of stationary and vehicle-mount types, as well as the supply from the grid. 
The community-owned facilities can include a hydrogen system composed of P2G and FC, a large stationary battery, and loads for common use. 


The power router is introduced to physically realize the variable connections between houses or houses and community-owned facilities. 
The main part of the power router is a switching circuit of a crossbar type as shown in \fref{fig:system_concept}. 
The connection relationship between multiple ports is determined by the on/off state of the switches. 
Power routing hereafter refers to the dynamic change in the relationship in accordance with supply and demand conditions. 


A preliminary version of the power router of line-switching type was developed and experimentally verified in \cite{Takuno.etal-2011b}.
However, the configuration needs to be enhanced for the purpose of the present study. 
Since the target of the original study was to switch the lines between sources and load, the router configuration was intended for unidirectional power flow. 
In this paper, we present a power router with four ports, each of which can accommodate bidirectional power flow. 
This is essential because the houses in this study are prosumers, i.e., yield bidirectional power flow of both distributing and receiving. 

\begin{figure}[tbp]
    \centering
    \includegraphics[width=7cm]{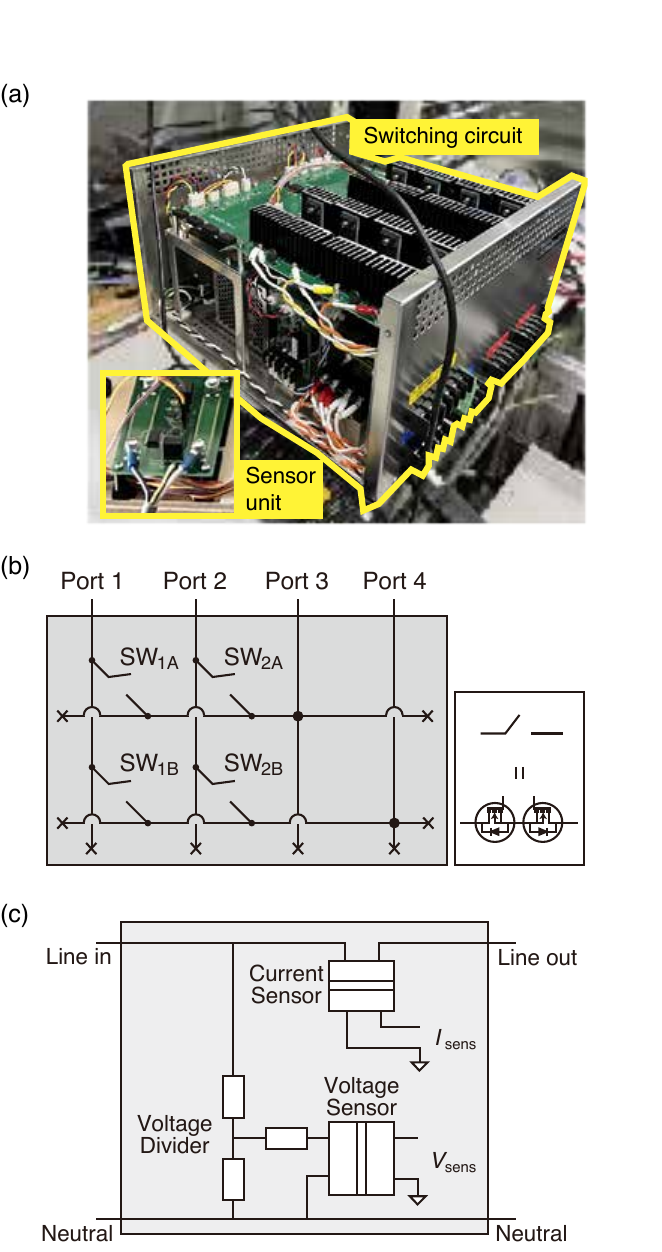}
    \caption{Prototype power router. (a) Photo of overview. (b) Switching circuit. (c) Circuit of sensor units.}
    \label{fig:setup_phys}
\end{figure}

\fref{fig:setup_phys}~(a) shows the overview of the developed router. 
The router has four ports, which are connected via a switching circuit. 
At each port, the voltage and current are measured by a sensor unit. 
The details are described below per component. 

\fref{fig:setup_phys}~(b) depicts the diagram of the switching circuit. 
Each switch unit is composed of two Metal-Oxide-Semiconductor Field-Effect Transistors (MOSFETs) in a back-to-back configuration for controlling the bidirectional power flow. 
We adopt silicon carbide MOSFETs of ratings $650\,\mathrm{V}$, $20\,\mathrm{A}$ (C3M0060065; Wolfspeed, Inc.).
The switching devices are selected for verification of the present demonstration; they should be replaced with ones of more capacity at the stage of social implementation. 


Note that the switching circuit is simplified from a full configuration of a four-port matrix. 
For a connection of arbitrary two pairs of two ports, the number of switches required is eight. 
The present router realizes a part of the full configuration that is enough for the verification in this paper. 

\fref{fig:setup_phys}~(c) depicts the circuit diagram of the sensor units. 
The diagram is simplified for visibility by omitting some components e.g. for ICs' powering and signal filtering. 
The line-to-neutral voltage is measured by a voltage divider and a galvanic isolation amplifier IC. 
The line current is measured by a Hall-effect-based current sensor IC. 
The sensor outputs are transferred to the controller board and can be monitored on a PC connected to the board in real time. 



The introduction of the power router addresses the aforementioned challenges in the following ways. 

\subsection{Self-consumption of Renewables}

The power routing enables the houses to share their excess generation of renewables and unused battery capacity with others in exchange for money. 
Generally, it is not realistic to install an extremely large stationary battery that can accommodate all the excess power from PV. 
Introducing an electric vehicle increases the capacity per house, but its availability depends on the use profile as a vehicle. 
Sharing these properties provides a better solution without increasing the capacity per house, taking into account the difference in demand and vehicle-use tendencies across houses \cite{Kalathil.etal-2019}. 
The inter-house power routing exploits the difference of the tendencies and matches houses of excess and short supplies. 

Still, the PV generation may exceed the total demand plus the batteries' capacity of the community in some cases such as daytime on sunny weekdays of mild temperatures. 
In this case, the excess power is routed to the community-owned battery or P2G. 
Based on the trade-off between conversion efficiency and energy density\cite{Evans.etal-2012}, the battery is used for short-term compensation of instantaneous and relatively small variation, whereas P2G for long-term and large-scale storage such as inter-seasonal one\cite{Fukaume.etal-2022}. 

Here, the physical distinguishment of power flow is essential for these sharing features. 
The strong tie between power and financial transactions requires the system to trace the power flow strictly to avoid the mismatch between the ledger and the real flow. 
Furthermore, regarding the stored energy in community-owned facilities such as P2G, the power-flow distinguishment at the power router is essential because it is impossible to tell who owns what portion of the total amount stored after it gets mixed. 

\subsection{Resilience}

Power routing also contributes to the enhancement of resilience. 
As mentioned above, the increase in self-consumption itself leads to enhanced resilience in terms of less dependency on the grid. 
Furthermore, the introduction of power routing provides a unique function in an emergency situation such as a loss of external supply. 
The power router can select the supply relationship so that electricity be delivered exclusively to loads with priority. 
For example, one can put priority on loads of public importance or urgency over other personal consumption. 
The power router ensures that critical loads are supplied regardless of the intention of each house involved in the system by physically establishing dedicated paths while blocking others. 
The line-switching routing realizes this power allocation in a direct way compared to a conventional way, such as calling for cooperation in saving electricity. 

\subsection{Traceability}

Third, the power routing enables us to track the origin of electricity for hydrogen production. 
As mentioned above, it is difficult to know how much portion of electricity comes from renewable sources in the conventional system. 
Since the power routing restricts the power source connected to P2G physically, it is possible to tell how much part of stored hydrogen is renewable, or to restrict hydrogen production only from fully renewable power. 

\section{Experimental Verification}

\subsection{Setups for Experiment}

In the following subsections, we confirm the operation of the developed router by experiments. 
An example of power routing is demonstrated with setups assuming a network of three houses. 
Each house is supplied not only by the commercial grid but also by its own power sources such as PV. 
Using their own supplies, the houses, as prosumers, carry out the bidirectional power transactions through the power router. 
In this sense, one of the houses can also be recognized as a combination of P2G and FC, corresponding to a load and a supply, respectively. 


Based on the above three-house system, we develop an experimental setup. 
In this study, we consider a combination of a DC power supply (ZX-1600HA; TAKASAGO, LTD.) and a power conditioning system (PCS) as a source installed in each house. 
We employ the PCS that is currently in commercial use for a household fuel cell in Japan. 
The supply from the commercial grid is modeled by a regulated AC supply (DP090RD; NF Corporation) in the constant-voltage mode set at $200\,\mathrm{Vrms}$, $60\,\mathrm{Hz}$. 
The load of each house is modeled by an electronic load. 

\begin{figure}[tbp]
    \centering
    \includegraphics[width=8.5cm]{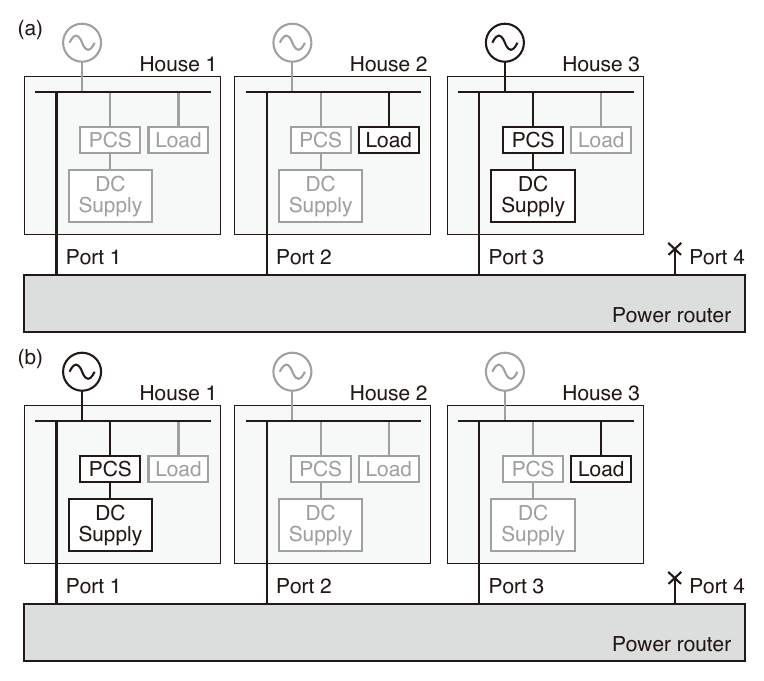}
    \caption{Experimental setups. (a) Mode A. (b) Mode B. }
    \label{fig:bidir_expr}
\end{figure}

We set two routing modes to confirm the bidirectional flow. 
\fref{fig:bidir_expr} depicts the configuration of the experimental verification. 
In mode A, house 3 is supposed to deliver its surplus power to house 2. 
House 1 is not involved in this exchange since the supply and demand are balanced within the house. 
Then in mode B, house 1 supplies house 3 with its excess generation. 
House 2 is not involved in this exchange for the same reason as house 1 in mode A. 
With these two modes, the bidirectional flow occurs at port 3. 

In the experimental setup, the supply and demand at each house are expressed only by a single device representing their sum.
For example, let us assume that the supply by the PCS is  $P_1$ and the consumption by the load is $P_2$ in a certain house.
When $P_1 > P_2$, the output of the PCS is set to $P_1 - P_2$ and no electronic load is installed in the house. 
When $P_1 < P_2$, the electronic load is set to $P_2 - P_1$ and the PCS is not installed. 
This is a simplification to demonstrate the scenario in this study with a limited number of devices available.
The DC supply at each house operates at a constant-power mode of $500\,\mathrm{W}$. 
The electronic loads operate at a constant-power mode of $500\,\mathrm{W}$. 

The two modes are altered by the switching circuit of the power router. 
The switching states of $(\mathrm{SW}_\mathrm{1A}$ , $\mathrm{SW}_\mathrm{1B}$ , $\mathrm{SW}_\mathrm{2A}$ , $\mathrm{SW}_\mathrm{2B})$ in \fref{fig:setup_phys}~(b) for modes A and B are $(\mathrm{OFF},\mathrm{OFF},\mathrm{ON},\mathrm{OFF})$, $(\mathrm{ON},\mathrm{OFF},\mathrm{OFF},\mathrm{OFF})$, respectively. 
It should be noted that the combination of switch states in the table is one of the possible candidates. 



\subsection{Results and discussions}



\begin{figure*}[tbp]
    \centerline{\includegraphics[width=15.5cm]{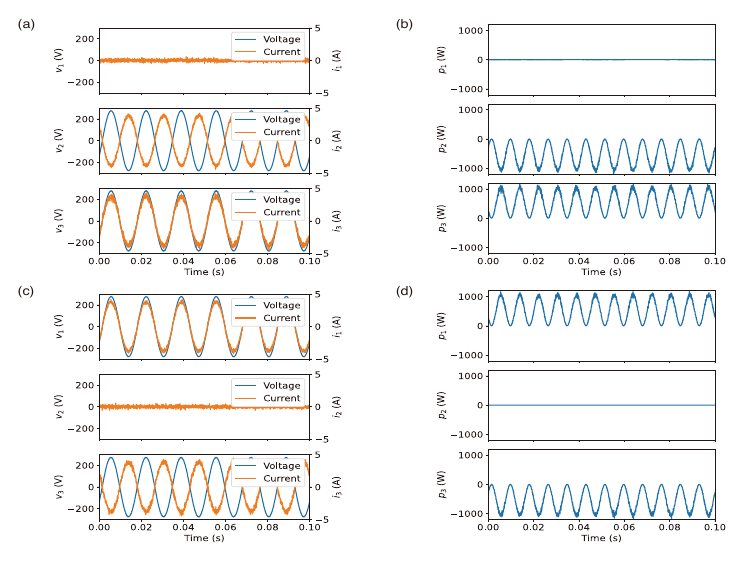}}
    \caption{Results of the experiments. (a) Voltage and current of mode A. (b) Power of mode A. (c) Voltage and current of mode B. (d) Power of mode B.}
    \label{fig:result}
\end{figure*}

\fref{fig:result}~(a) shows the measurements of voltage and current on the sensor units in mode A. 
The current was measured with the direction of inflow to the router (namely, outflow from the house) as positive. 
The voltage and current were in-phase at port 3 and anti-phase at port 2. This is because house 3 supplies power to house 2. 
There was no power flow at port 1 because house 1 had no power transaction with other houses. 
These can also be confirmed with a power calculation by multiplying the voltage and current measurements.
\fref{fig:result}~(b) shows the instantaneous power flow in mode A. 
The direction of power flow coincided with the supply relationship between the houses. 
The average power is approximately $500\,\mathrm{W}$, which is also consistent with the experimental setup. 
Now note that the fluctuation observed in the power waveforms is due not to the introduction of power routers but to the imperfect tuning of the embedded sensor circuits, which we will address in future development. 

Then we move on to the results of mode B. 
\fref{fig:result}~(c) shows the voltage and current measurements in mode B. 
The voltage and current were in-phase at port 1 and anti-phase at port 3, while no current flowed at port 2. 
As a result, the instantaneous power flow can be calculated as shown in \fref{fig:result}~(d). 
The sign of the power flow changed following the alternation of the supply relationship. 



The comparison of \fref{fig:result}~(b) with \fref{fig:result}~(d) indicates the bidirectional power flow at port 3. 

\section{Conclusions}
In this paper, we proposed a system of bidirectional power routing for inter-house multi-energy management. 
The proposal exploits the concept of sharing economy to address the urgent issues in power systems while offering a reasonable motivation for households involved. 
The proposed system was also experimentally demonstrated in an example scenario. 
The results support the feasibility of the proposed energy management with the power router. 

The next step would be software development to determine the power routing table at each unit time according to the weather forecast, consumption expectation, and so on. 

\section*{Acknowledgment}

The authors thank Professor Takashi Hikihara of Kyoto University for his invaluable advice. 

\bibliographystyle{IEEEtran}
\bibliography{IECON2024}

\end{document}